\documentclass[twocolumn,prd,nofootinbib,aps,floats,floatfix,amsmath,amssymb,longbibliography,
secnumarabic]{revtex4} 
\usepackage[utf8]{inputenc}
\usepackage{graphicx}
\usepackage{amsmath}
\usepackage{amssymb}
\usepackage[a4paper, total={6.5in, 10in}]{geometry}
\usepackage{hyperref}
\usepackage{mathtools}
\usepackage{import}
\usepackage{mathtools}
\usepackage{accents}

\hypersetup{
    colorlinks=true,
    linkcolor=blue(ryb),
    filecolor=blue(ryb), 
    citecolor=blue(ryb),
    urlcolor=blue(ryb)}
\usepackage{slashed}

\newcommand{\be}{\begin{equation}}
\newcommand{\ee}{\end{equation}}
\newcommand{\ba}{\begin{array}}
\newcommand{\ea}{\end{array}}
\newcommand{\bea}{\begin{eqnarray}}
\newcommand{\eea}{\end{eqnarray}}

\newcommand{\nn}{\nonumber}

\begin{document}

\title{Constraints on Heavy Neutral Leptons interacting with a singlet scalar}
\author{James M.\ Cline}
\affiliation{McGill University, Department of Physics, 3600 University St.,
Montr\'eal, QC H3A2T8 Canada}
\author{Guillermo Gambini}
\affiliation{McGill University, Department of Physics, 3600 University St.,
Montr\'eal, QC H3A2T8 Canada}
\affiliation{Instituto de F\'isica Gleb Wataghin, UNICAMP, Rua S\'ergio Buarque de Holanda 777, 13083-859, Campinas-SP, Brasil}

\begin{abstract}
Heavy neutral leptons (HNLs) are an attractive minimal extension of the Standard Model, as is a singlet scalar $s$ mixing with the Higgs boson.  If both are present, it is natural for HNLs to interact with $s$.  For a light singlet,
the decay $N\to s\nu$ can dominate over weak HNL decays.  We 
reinterpret existing constraints on HNL mixing from the DELPHI, CHARM and Belle experiments for 0.5-100 GeV mass HNLs, taking into account the new decay channel. Although the constraints are typically weakened, in some cases they can become stronger, due to observable $s\to\ell^+\ell^-$ decays in the detectors. The method presented here could be used to recast constraints from other (older) experiments without resorting to computationally expensive Monte Carlo simulations. In addition, we update and correct some errors in the analysis of the original constraints, in the absence of the singlet.
\end{abstract}

\definecolor{blue(ryb)}{rgb}{0.01, 0.28, 1.0}
\maketitle

\section{Introduction}
Right-handed neutrinos $N_i$ are the minimal extension of the standard model (SM) needed to explain neutrino masses.  Although their masses might be expected to far exceed the weak scale, in principle there is no restriction on how light they could be, down to the scale $m_\nu$ of active neutrino masses.  For $N$ masses in the range $0.1-1000\,$GeV, it is common to refer to them as heavy neutral leptons (HNLs).  They
mix with light neutrino flavors $\nu_i$ with mixing angles of order $U\sim \sqrt{m_\nu/m_N}$. The phenomenological implications of HNLs are therefore enhanced if they are relatively light \cite{Cvetic:2013eza,Cvetic:2014nla,Cvetic:2020lyh}.  They can be produced in 
$e^+ e^-$ collisions \cite{Dittmar:1989yg,L3:1992xaz,DELPHI:1996qcc,L3:2001xsz,L3:2001zfe}, neutrino beams \cite{Pais:1975dg,NuTeV:1999kej,Asaka:2012bb,MicroBooNE:2019izn,Breitbach:2021gvv},
in beam dump experiments (from the 
decays of mesons) \cite{Gronau:1984ct,Gninenko:2012anz,Bonivento:2013jag,Drewes:2018gkc}, and in the early Universe, leading to constraints from Big Bang Nucleosynthesis \cite{Dolgov:2000jw,Sabti:2020yrt,Boyarsky:2020dzc,Bondarenko:2021cpc}. HNL oscillations could be observed in $W$, $\tau$, and $B_c$ rare decays  \cite{Cvetic:2018elt,Tapia:2019coy,Tapia:2021gne} and its nature (Dirac or Majorana) could be inferred from rare meson decays \cite{Cvetic:2010rw,Cvetic:2015naa} 
or Z boson decays \cite{Blondel:2021mss}. A sufficiently weakly coupled HNL can be a viable cold dark matter candidate
\cite{Asaka:2005an,Asaka:2005pn,Boyarsky:2009ix,Cline:2020mdt}.
HNLs have been constrained by 
searches at the Large Hadron Collider \cite{CMS:2018iaf,ATLAS:2019kpx}.

Another simple and highly motivated extension of the SM is
a singlet scalar field $s$ that mixes with the Higgs boson
through the interaction $|H|^2s^2$, if $s$ gets a vacuum expectation value.  
Such singlets are constrained by collider searches \cite{acciarri1996search, cms2012search, LHCb:2018cjc} and Higgs decays \cite{Falkowski:2015iwa,Bojarski:2015kra}, and their possible enhancement of the electroweak phase transition in the early universe could produce observable gravitational waves \cite{Leitao:2015fmj,Huang:2016cjm,Hashino:2016xoj,Vaskonen:2016yiu,Cline:2021iff}, and facilitate electroweak baryogenesis \cite{Anderson:1991zb,McDonald:1993ey,Choi:1993cv,Espinosa:2011ax,Cline:2012hg}.   In the present work we will be interested in relatively light singlets.
Recent constraints are summarized in Ref.\ \cite{Winkler:2018qyg}, for example.

If both HNLs and singlets exist in nature, they can interact with each other via the Lagrangian term \cite{Hostert:2020xku,deGouvea:2019qre,Sanchez-Vega:2014rka,Alvarez-Salazar:2019cxw}
\be
    {\cal L} \ni g_s s\bar N N\,,
\ee
which is possible both for Dirac or Majorana HNLs.
This scenario was proposed in Ref.\ \cite{Cline:2020mdt} to enable a species of HNLs to be dark matter, with a thermal relic density from annihilations
$N\bar N \to s s$, or $N\bar N\to s^*\to f\bar f$, where
$f$ is any SM particle that couples to the Higgs boson.
In a generic theory of HNLs that mix with light neutrinos,
the mixing would generate the operator $s\bar N\nu$, opening the new decay channel $N\to \nu s$ if $m_s < m_N$.  In this case, the usual limits on the $N$-$\nu$ mixing angle $U$ will be modified, relative to the usual assumption that $N$ decays only through the weak interactions.  In this way, regions of parameter space in the $m_N$-$U$ plane, that are normally considered to be ruled out, could be reopened---or in some cases constraints can become stronger, as we will show.
It is the purpose of this paper to estimate how the modified constraints vary with the singlet mass $m_s$, its mixing
with the Higgs $\theta_s$, and the coupling $g_s$.  The possible values of
$g_s$ are constrained in the special scenario of Ref.\ \cite{Cline:2020mdt}
where they determine the dark matter relic density.  In this study we take a more generic approach and consider $g_s$ to be a free parameter, since it is not otherwise constrained.

This is a challenging task, since it requires the reinterpretation of experimental limits that must be dealt with individually for each experiment.  We therefore choose to limit our investigation to the HNL mass range $0.3-100$\,GeV, where current constraints are set by three experiments: DELPHI (from LEP) \cite{DELPHI:1996qcc}, CHARM \cite{charm}, and Belle \cite{liventsev}\footnote{For lighter and heavier HNLs see, for example, \cite{Arguelles:2021dqn} and \cite{Das:2015toa,Das:2016hof}, respectively. Limits from ongoing and future experiments can be found here \cite{Chun:2019nwi}.}. Despite the fact that ATLAS \cite{ATLAS:2019kpx} and CMS \cite{CMS:2018iaf} set competitive limits for $|U_e|^2$ and $|U_\mu|^2$, these are not much stronger than DELPHI on our region of interest (below 10 GeV); hence we focus on the first three experiments. 

Although we have tried to be as quantitative as possible, our results should be considered as indicative of more definitive
limits that would require a dedicated reanalysis of data for each experiment (as opposed to recasting published limits), which is beyond the scope of this study.

As a first step, we must be able to reproduce existing constraints in the absence of the singlet coupling.  This part of the exercise revealed that published limits from 
CHARM and Belle change somewhat when updated branching fractions for HNL production are employed, or other corrections that we describe in the main body of the paper.  Thus another result from the present work is  improved limits from these
experiments, even in the absence of a scalar singlet.

In section \ref{secII} we will present the main results for the three experiments, describing the essential characteristics relevant to our study of each one.  Further details are relegated to the appendices. In section \ref{secIII} we put our results into the perspective of independent 
constraints on the singlet scalar, to show the relation of those limits
to the recasted ones derived in section \ref{secII}. Conclusions are given in section \ref{secIV}.

\begin{figure*}[t]
    \centering
    \includegraphics[scale=0.75]{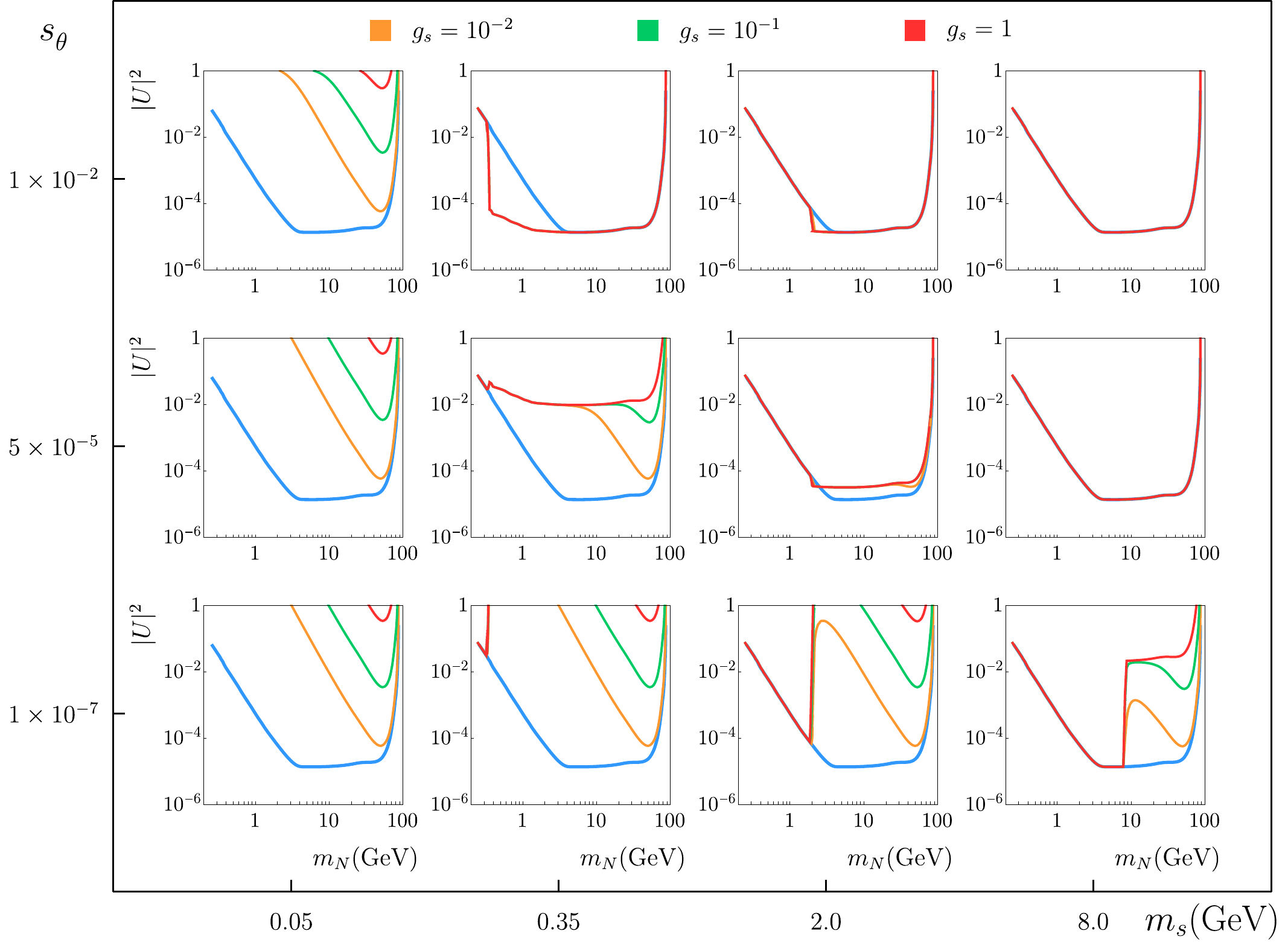}
    \caption{DELPHI: Upper limits on $|U|^2$ for heavy neutral leptons that couple to a light singlet scalar, which mixes with the Higgs boson, in the HNL mass range from $2\,m_\mu$ up to the kinematical limit $m_Z$ at LEP. The light blue contour is DELPHI's limit on $|U|^2$ at the $95\%$ C.L.\ in the case of no singlet,
    which is independent on the HNL flavor (see figure \ref{results0}).}
    \label{fig:DELPHI}
\end{figure*}

\section{Recasted constraints on HNL$-$active neutrino mixings}
\label{secII}

Our goal is to arrive at a reasonable approximation to how HNL mixing constraints are changed by $N\to s\nu$ decays, without however doing a full reanalysis of experimental data, which in any case would not be feasible given our limited understanding of detector responses and systematic errors, and lack of access to the data.  Instead we will
theoretically compute the number of events expected to be produced in a given experiment, as a function of $m_N$ and
mixing $U$, and use the existing constraints to calibrate the detection efficiency.
As a preview, we present our new limits from the DELPHI experiment in Fig.\ \ref{fig:DELPHI}, discussed in more detail below, where the original constraint is the light blue contour and modified ones depending upon the singlet mass $m_s$, its mixing with the Higgs $\theta$ and its coupling $g_s$ to the HNL are shown.

\begin{figure*}
    \centering
    \includegraphics[scale=0.85]{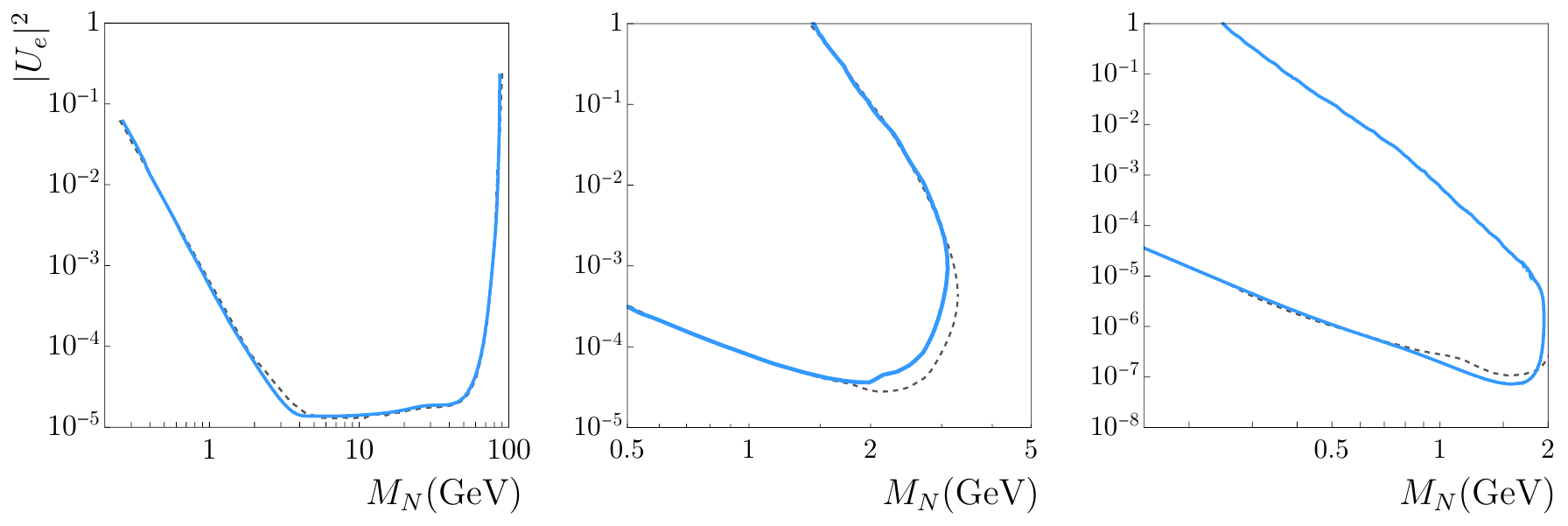}
    \caption{\textbf{(Left)} Dashed curve: DELPHI limits at the 95\% confidence level on $|U|^2$ as a function of the HNL mass \cite{DELPHI:1996qcc}. These limits apply equally to all neutrino flavors. Light blue curve: our result using equation \ref{eq-frac}. \textbf{(Center)} Dashed curve: Belle limits at 90$\%$ C.L. on $|U_e|^2$ \cite{liventsev}. Light blue curve: our result using the procedure described above. The difference in the region $\gtrsim$ 2 GeV is due to the use of updated formulas for HNL production and decay \citep{Bondarenko:2018ptm}, and the treatment of HNL momenta (see appendix). \textbf{(Right)} Dashed curve: CHARM limits at 90$\%$ C.L. on $|U_e|^2$ \cite{charm}. Light blue curve: our result for these limits. A similar result is obtained in \cite{Boiarska:2021yho}.}
    \label{results0}
\end{figure*}

To obtain these results, we first compute the number $N_N$ of HNLs produced in $Z\to N\nu$ decays at LEP, as a function of $m_N$ and $U$, relative to the total number of $N_Z$ of $Z$ bosons.  A branching fraction $f = N_N/N_Z$ is excluded 
for $f$ greater than some value $f_0$ depending upon the
detection efficiency.  By varying $f_0$, we can produce contours in the $m_N$-$U$ plane.  If one of those contours matches the existing DELPHI limit, we can adopt the corresponding $f_0$ value and claim to sufficiently understand how to reproduce the original constraint, and then investigate how it changes in the presence of the new decay channel.  The effectiveness of this strategy is illustrated in Fig.\ \ref{results0} (left), which shows the good agreement between the original LEP limit (dashed) and our reconstruction (solid).  (For the other two experiments, where the agreement seems less good, we will argue below that our reconstructions more accurately reflect what the true limits should be.)

A possible effect of $N\to s\nu$ is that the HNL decays invisibly, since $s$ may be too long-lived to decay within the detector.  The constraint on $U$ then gets weakened according to the branching fraction for weak HNL decays versus the $N\to s\nu$ channel.  On the other hand, if $s$ is short-lived and
decays into $e^+e^-$ or $\mu^+\mu^-$, within the detector, those final states might mimic weak decays of $N$, leading to new constrained regions of parameter space, or they might be rejected by the search, depending on the experiment.  In the
following, we will assume that electromagnetic decays of $s$ could have mimicked weak decays of $N$ for DELPHI and CHARM,
but not for Belle, as will be explained.  Therefore in some regions of parameter space, $N\to s\nu$ can give rise to new excluded regions for DELPHI and CHARM,
while for Belle it can only relax the existing constraints.

\begin{figure}[t]
    \centering
    \includegraphics[scale=0.2]{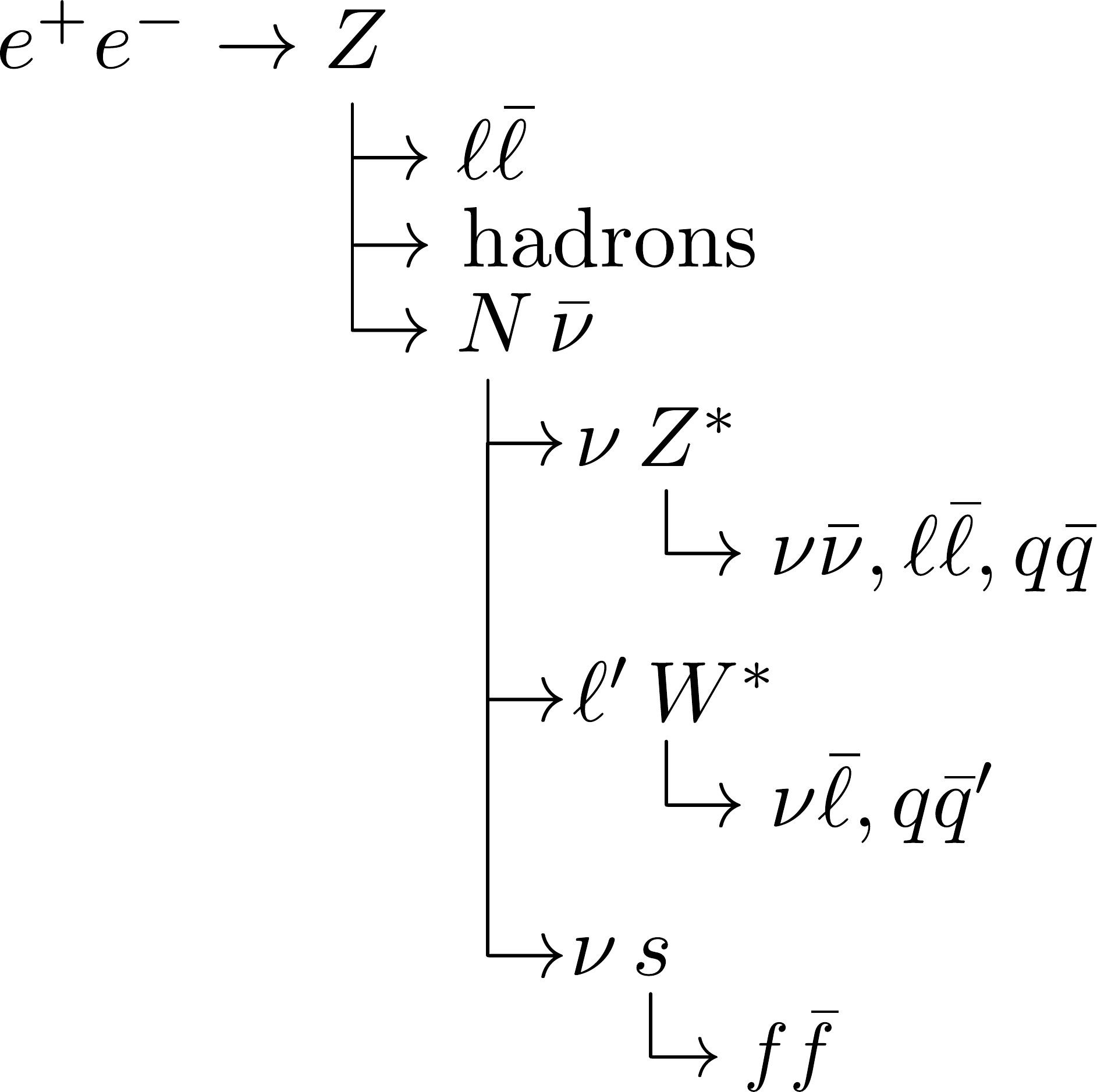}
    \caption{Possible decays of the $Z$ boson including HNL intermediate states.}
    \label{Zchain}
\end{figure}

\subsection{DELPHI}

The DELPHI detector at LEP I collected $3.3 \times 10^6$ hadronic $Z$ decays from 1991 to 1994 \cite{DELPHI:1996qcc}.
In these decays, HNLs $N$ could be produced
via $Z\rightarrow N \bar{\nu}$ and $Z\rightarrow \bar{N} \nu$ through the mixing with light neutrinos 
\bea
    N' &\cong& N +  U_{\alpha}\nu_\alpha\nn\\
    \nu'_\alpha &\cong& \nu_\alpha - U_\alpha N
\eea
in transforming  between the Lagrangian and mass eigenstates. The branching ratio is given by
\bea
    \text{BR}(Z\rightarrow N \bar{\nu}_\alpha) &=& \text{BR}(Z \rightarrow \nu_\alpha \bar{\nu}_\alpha) |U_\alpha|^2 \\
    &\times& \left( 1- \frac{m_N^2}{M_Z^2} \right)^2 \left( 1+ \frac{1}{2} \frac{m_N^2}{M_Z^2} \right)\,,\nn
    \label{eq1DELPHI}
\eea
where $\text{BR}(Z \rightarrow \nu_\alpha \bar{\nu}_\alpha) \approx 0.063$ for any $\alpha=e,\mu,\tau$.\footnote{Unlike the other experiments, LEP limits apply equally to all flavors of HNLs \cite{Dittmar:1989yg}.}\ \ The mean decay length of the HNLs is $L \cong {3}|U|^{-2}({\text{GeV}}/{m_N})^6\ \text{cm}.$
The DELPHI Collaboration studied three different decay topologies: $\nu \ell \bar{\ell}$, $\nu q\bar{q}$, and $\ell q \bar{q}'$ where $\ell=e,\mu,\tau$, $q=u,d,s,c,b$, and $q\bar{q}'=u\bar{d}, c\bar{s}$ plus charge conjugate states. These are illustrated in Fig.\ \ref{Zchain}.
The fraction of $Z$ bosons leading to observed HNLs decaying inside the detector (via weak interactions) 
is
\begin{equation}
    f_{w} = \text{BR}(Z\rightarrow N) \left( 1 - e^{-D_L/L}  \right)
    \varepsilon(m_N),
    \label{eq-frac}
\end{equation}
where $\text{BR}(Z\rightarrow N) = \text{BR}(Z\rightarrow N \bar{\nu}_\alpha) + \text{BR}(Z\rightarrow \bar{N} \nu_\alpha)$, and the reconstruction efficiency $\varepsilon(m_N)$ is taken from Fig.\ 4 of Ref.\ \cite{DELPHI:1996qcc}.  $D_L$ is the length of the region in which decays are observed.  As described in the appendix, we infer this parameter (obtaining $D_L=200\,$cm) when reconstructing the published limit for weak HNL decays.

Four searches were performed covering HNL masses from $2\,m_\mu$ up to the kinematic limit $m_Z$: (i) the decay products of HNLs with short lifetimes and small masses (monojets) were searched for within 12\,cm of the interaction point;
(ii) hadronic systems (acoplanar and acollinear jets) were produced by HNLs with short lifetimes and large masses (40-80 GeV); (iii)  a HNL with an intermediate lifetime  (decaying at radii from 12 to 110 cm) could have produced an isolated set of charged particle tracks originating from the same vertex; (iv) HNLs with long lifetimes would decay in the detection region where charged particle tracks cannot be reconstructed (110 to 300 cm), so this search had to rely on localized clusters of energy depositions and hits in the outermost layers of the detector. DELPHI computed the detection efficiencies from $10^5$ signal events for HNL masses from 1.5 to 85 GeV and mean decay lengths from 0 to 2000 cm.

In our study, we calculated the fraction of HNLs decaying inside the DELPHI detector considering detector length $D_L=$ 200 cm and using the global efficiency shown in Fig.\ 4 of Ref.\ \cite{DELPHI:1996qcc}. We scaled our result to match the published DELPHI constraint; see figure \ref{results0}, left.

The next step is to add the effects of the 
gauge singlet scalar $s$ that couples to HNLs with strength $g_s$ and mixes with the Higgs boson through a small angle $\theta$,
\begin{equation}
    \mathcal{L} \ni - g_s s\bar{N}N - \frac{s_\theta \, m_f}{v} s \bar{f}f\,, 
    \label{eq-frac2}
\end{equation}
where $s_\theta=\sin\theta$, $v=174\,$GeV is the complex Higgs vacuum expectation value (VEV), and $f$ represents SM fermions with mass $m_f$. The final states listed in
Fig.\ \ref{Zchain} are such that 
decays of the singlet scalar into fermions ($s \rightarrow f \bar{f}$) will only affect the signals for the first two event candidates, {\it i.e.,}  $\nu \ell \bar{\ell}$ and $\nu q\bar{q}$  (see figure \ref{Zchain}). However, the total decay width of the HNL will increase and so will the probability for the HNL to decay inside the detector. Consequently, we replace Eq.\ (\ref{eq-frac}) with the fraction that includes the additional events from $s\to f\bar f$ decays,
\bea
    f_{w+s} &=&  \text{BR}(Z\rightarrow N) \Big[ \text{BR}_w \times (1 - e^{-D_L/L_N})\nn\\
    &+&\text{BR}_s \times (1-e^{-D_L/(L_N+L_s)})  \Big]\,,
    \label{eq-frac-ws}
\eea
where 
\bea
    \text{BR}_w &=& \frac{\Gamma_N(\text{weak decays})}{\Gamma_N(\text{weak decays})+\Gamma(N\rightarrow s \nu)}\,,\nn\\
    \text{BR}_s &=& 1-\text{BR}_w\,,
\eea
and $L_s$, the mean decay length of the gauge singlet scalar $s$, is calculated following Refs.\ \cite{Fradette:2017sdd, Winkler:2018qyg}.  The corresponding decay length $L_N$ of the HNL is given in the Appendix. We note that HNLs can also be produced in $Z$ boson 3-body decays $e^+e^- \to Z \to s Z^* \to s N \bar{\nu}$ \cite{acciarri1996search}, but these are suppressed by $s_\theta^2$ and kinematical factors in comparison with $Z \to N \bar{\nu}$ and $Z \to N \nu$, so this production channel is not shown in Fig.\,\ref{Zchain} and, consequently, $\text{BR}(Z\rightarrow N)$ remains the same as in Eq. (\ref{eq-frac}). With these modifications, we obtain the upper bounds for HNL-active neutrino mixing shown in Fig.\ \ref{fig:DELPHI} (left).  

The modified limits on $|U|^2$ can be understood as the result of the competition between HNL weak (3-body) and scalar (2-body) decays, through the factors BR$_w$ and BR$_s$, and the interplay between the altered decay length of the HNL $L_{N}(\text{weak}+s)$ and that of $s$. For example, the bottom right plot of Fig. \ref{fig:DELPHI} shows the weakened limits  starting at the kinematic threshold  $m_N>m_s =8$\,GeV.  As the coupling $g_s$ between $N$ and $s$ increases,
$N\to s\nu$ decays become more prevalent.  These are invisible decays at small mixing $\theta = 10^{-7}$,
decreasing the number of signal events and weakening the limit on $|U|^2$.  On the other hand,
at larger singlet-Higgs mixing $\theta \ge 5\times 10^{-5}$,
the singlet decays to $f\bar f$ with a short enough decay length for the final state particles to be observed, as though they were coming from weak decays.  We assume that experimental sensitivity to these events is similar to that for the weak decays, resulting in an unmodified limit relative to the published result.

As we move to the left in Fig.\ \ref{fig:DELPHI}, looking at the columns corresponding to smaller values of $m_s$, the kinematic threshold discontinuity for $N\to s\nu$ also moves to the left, until the first column where it is no longer visible since $m_s < m_N$ for the range of $m_N$ considered.
The pattern described for the right-most column is similar, 
except that the decay length $L_s$ is additionally increased by the small $m_s$ suppressing $s\to f\bar f$ decays, leading to more invisible decays and generally weaker limits.

Exceptionally, there are several regions where the constraint on $|U|^2$ is strengthened, most notably in the plot where $m_s=350\,$MeV, 
$\theta = 10^{-2}$, $m_s<m_N \lesssim 4\,$GeV.  It can be understood through the increased signal from $N\to s\nu$
followed by $s\to f\bar f$ compared to weak decays.
This excluded region eventually merges back to the pure weak decay limit as $m_N$ increases, since the weak decay rate scales as $m_N^5$, while the two-body rate scales as $m_N^3$.

\begin{figure}[t]
    \centering
   \includegraphics[scale=0.35]{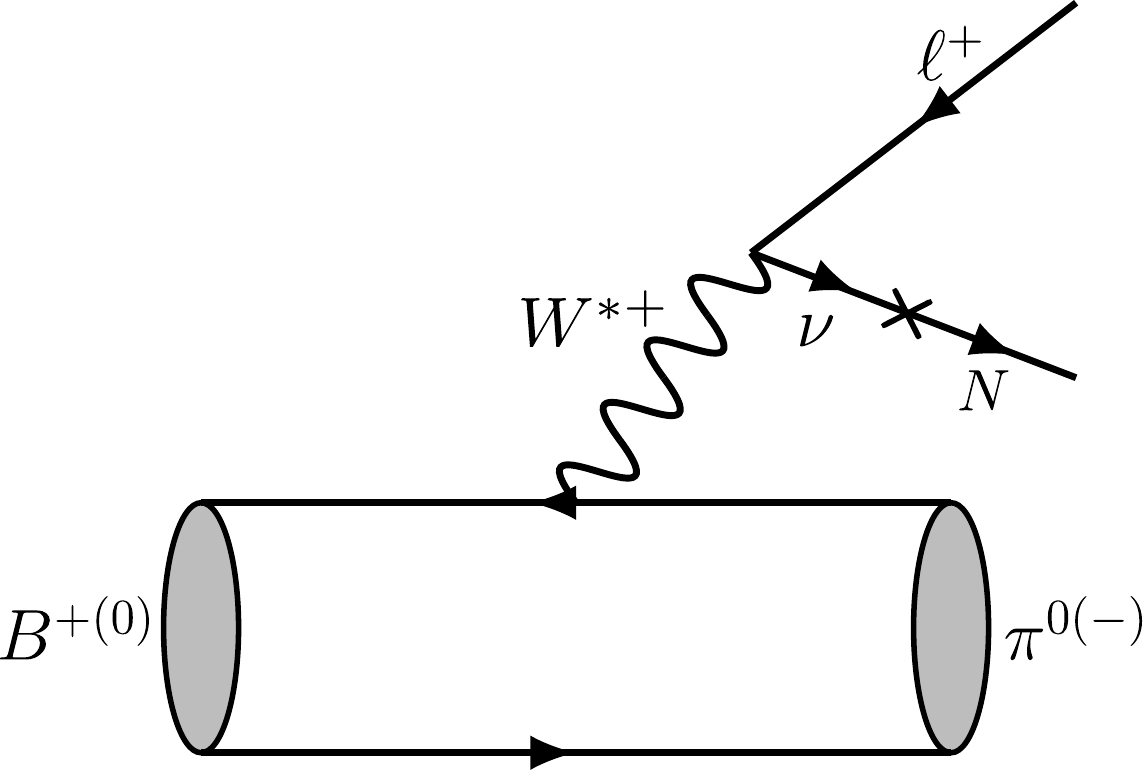}
    \caption{Heavy neutral lepton production from $B$ meson decays.}
    \label{fig:Belle}
\end{figure}

\begin{figure*}[t]
    \centering
    \includegraphics[scale=0.65]{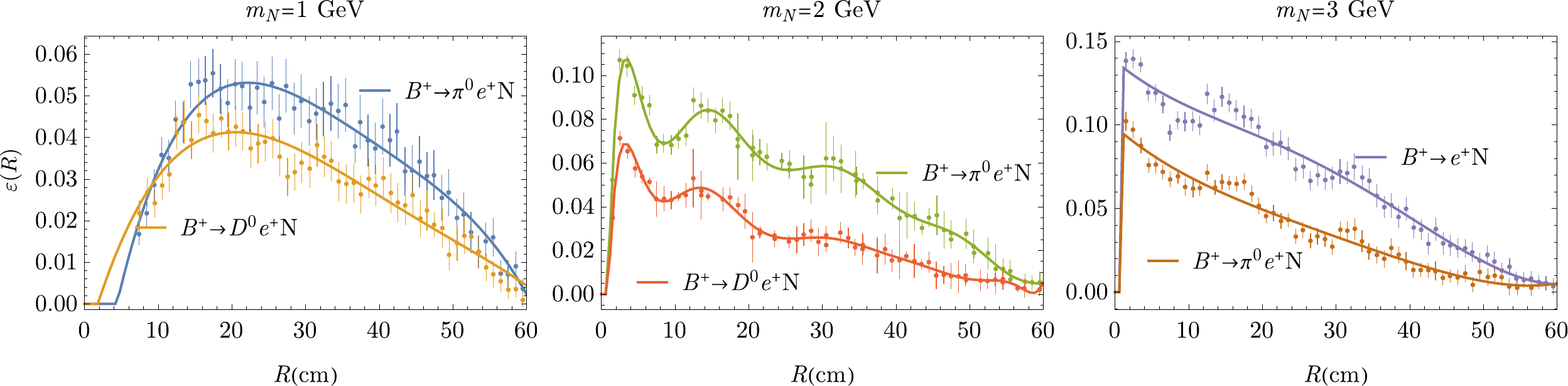}
    \caption{Belle efficiencies $\varepsilon (R)$ for different HNL production modes. From left to right $m_N=$ 1,2,3 GeV, respectively. Solid curves: our fits for the efficiencies.  To interpolate between different values of masses, we use the 
    mass-dependent efficiency curves shown in Fig.\ 2 of Ref.\ \cite{liventsev}.}
    \label{Belle-eff}
\end{figure*}

\begin{figure*}[t]
    \centering
    \includegraphics[scale=0.8]{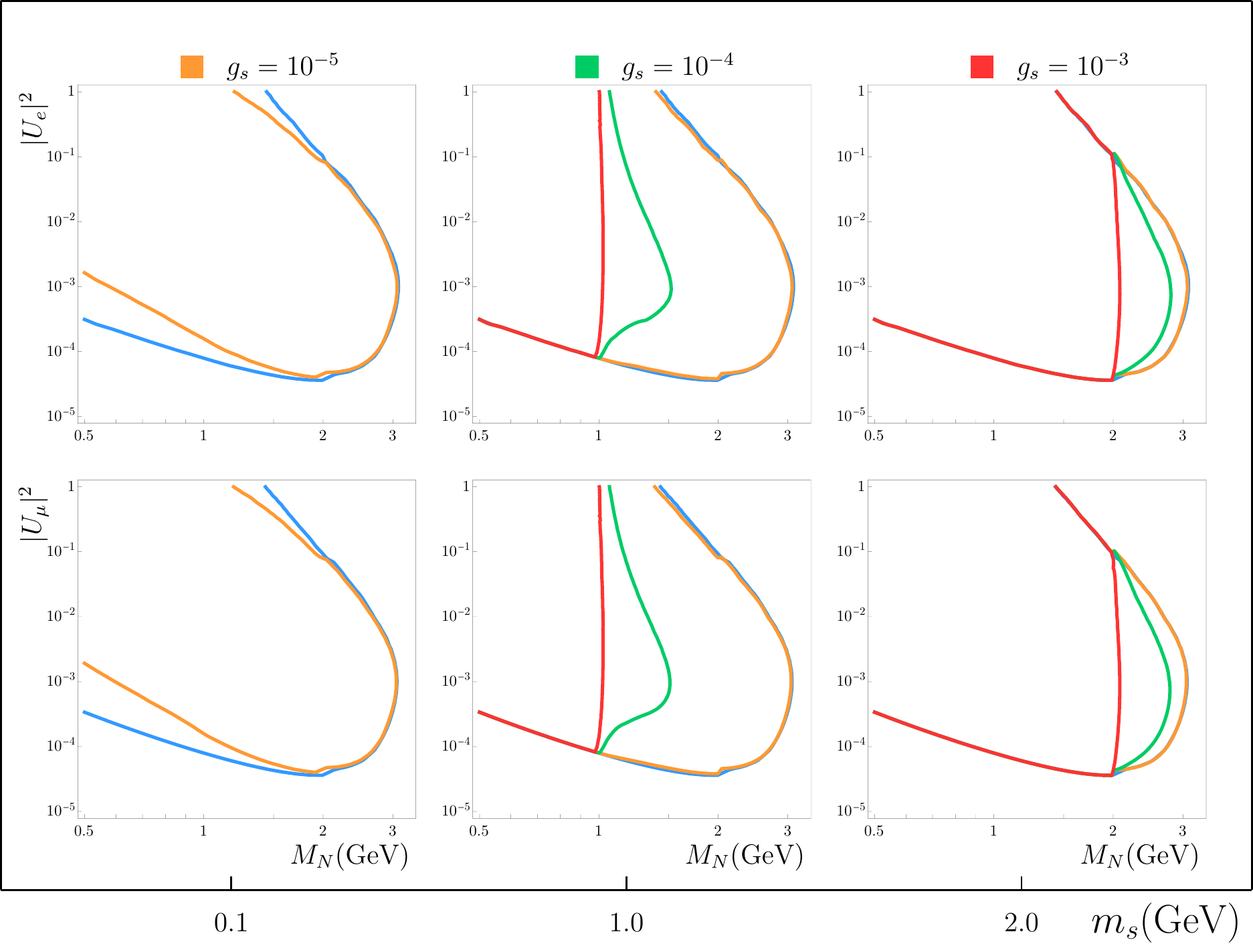}
   \caption{Belle: Light blue curves show our reconstruction of the Belle limits on weak decays of HNLs.  Other colors show the relaxed constraints for different values of the coupling $g_s$ due to $N\to s\nu$
   decays.}
   \label{Belle1}
\end{figure*}

\subsection{Belle}

The Belle experiment searched for direct HNL decays ($N \rightarrow \ell^\pm \pi^\mp$, $\ell=e,\mu$) at the KEKB $e^+e^-$ collider, where $B\bar{B}$ pairs created at the $\Upsilon (4S)$ resonance in $e^+e^-$ collisions.  The HNL production mechanism from $B$ decays is illustrated in figure \ref{fig:Belle}.
The number of detected heavy neutral leptons ($N_N$) is given by \cite{liventsev}
\bea
    N_N &=& 2 N_{B\bar{B}}\, \text{BR}(B \rightarrow N) \, \text{BR}(N \rightarrow \ell \pi)\nn\\
    &\times&\int \frac{m_N\Gamma}{p_N} \exp\left(-\frac{m_N\Gamma R}{p_N} \right) \, \varepsilon (R)\, dR\,,
    \label{BelleEq}
\eea
where $p_N$ is the momentum of the HNL, $\Gamma$ is its total decay width, and $\varepsilon(R)$ is its reconstruction efficiency as a function of the distance $R$ from the interaction point.  $\varepsilon(R)$ is shown for three values of $m_N$ and the most relevant $B$ decay channels in Fig.\ \ref{Belle-eff}.

The most favorable mass range in which to look for HNLs at Belle is $M_K <m_N<M_B$. For this reason, the total branching fraction for HNL production
\begin{equation}
\text{BR}(B \rightarrow N) = \sum_X  \text{BR}(B \rightarrow X \ell_1 N),   
\end{equation}
includes $X = \pi, \eta, \rho, \omega, \eta' ,\phi, D, D^*$ for semileptonic decays and
\begin{equation}
\text{BR}(B \rightarrow N) =  \text{BR}(B \rightarrow \ell_1 N),   
\end{equation}
for purely leptonic decays \cite{liventsev}. 
Assuming HNLs to be Majorana,\footnote{For Dirac HNLs, only the processes where the signal fermion $\ell_2$ (produced by the HNL) has the opposite electric charge with respect to the production fermion $\ell_1$ (coming from the decay of the $B$ meson) should be considered.}   

\bea
    \text{BR}(N\rightarrow \ell_2 \pi) &=& \text{BR}(N\rightarrow \ell_2^- \pi^+)\nn\\ &+& \text{BR}(N\rightarrow \ell_2^+ \pi^-)\,,
    \label{BrN}
\eea
where $\text{BR}(N\rightarrow \ell_2^+ \pi^-)=\text{BR}(N\rightarrow \ell_2^- \pi^+)$. Therefore the signal events of the form  $\ell_1 \ell_2 \pi$ are $e^+e^+\pi^-,$ $e^+e^-\pi^+,$ $e^{-}e^{+}\pi^{-},$ and $e^{-}e^{-}\pi^{+}$.
For the calculations of the HNL production and decay products we followed Ref.\  \cite{Bondarenko:2018ptm},\footnote{We used $M_{\text{pole}}=\infty$ \cite{FlavourLatticeAveragingGroup:2019iem,DeVries:2020jbs} instead of $5.65$ GeV \cite{Bondarenko:2018ptm}  for $f_0^{B\rightarrow \pi}$ in the calculations of the $B$ meson form factors.}\ which is updated relative to the values used in the Belle analysis, and leads to some differences in our determination of the standard HNL constraint compared to the published version.

We start by considering the standard assumption of weak decays only. At large masses, the decay length $L_N$ is so short that the HNLs decay close to the interaction point, where the reconstruction efficiencies at short distances $R\cong 0$ are negligible (see figure \ref{Belle-eff}), and no constraint
on $|U|^2$ arises.
At lower masses, and for sufficiently small mixing,  HNLs decay outside of the detector, again leading to no constraint. On the other hand, if $|U|^2$ is sufficiently large, the mean HNL decay length can be so small that their decays do not meet experimental selection criteria, similarly to the case of heavy HNLs.  Therefore constraints arise only for relatively small $m_N$ and an intermediate range of $|U|^2$, as shown in Fig.\ \ref{results0} (center).
The difference between our reconstructed limit (solid) and the original one (dashed) is due to the updated branching ratios mentioned above.

Like for the DELPHI search, the $N$-$s$ interaction  introduces competition between three-body weak decays and two-body scalar decays of the HNL.
But in contrast to DELPHI, at Belle the $N\rightarrow s \nu$ decays cannot contribute to signal events, which are taken to be pions and
charged leptons in the final state.  Although $s$ could decay to
$\ell^+\ell^-$ or $\pi\pi$, it can never produce the combination
$\pi\ell$ which is required by the search.  Therefore $N\to s\nu$
is invisible in the Belle analysis, and these decays can only weaken the limit on $|U|^2$, independently of the size of the singlet-Higgs mixing angle $\theta$.

Fig.\ \ref{Belle1} shows our results for three choices of $m_s$ and a range of values for $g_s$, displaying again the kinematic threshold discontinuity whenever $m_N = m_s$.  As stated, the effect of the invisible decays is only to weaken the bounds.  A borderline case is
the coupling $g_s=10^{-5}$ (green lines), where at high HNL masses there is no change relative to the purely weak decay bounds, since the weak decays dominate at large $m_N$: their rate scales as $m_N^5$, while the rate for $N\to s\nu$ goes as $m_N^3$.  At lower $m_N$, provided that
$m_s < m_N$ (left column), one observes a weakening of the limits.
On the other hand, for large enough $g_s$, the limits can disappear entirely, again provided that $m_s < m_N$.

\subsection{CHARM}

\begin{figure*}[t]
    \centering
    \includegraphics[scale=0.75]{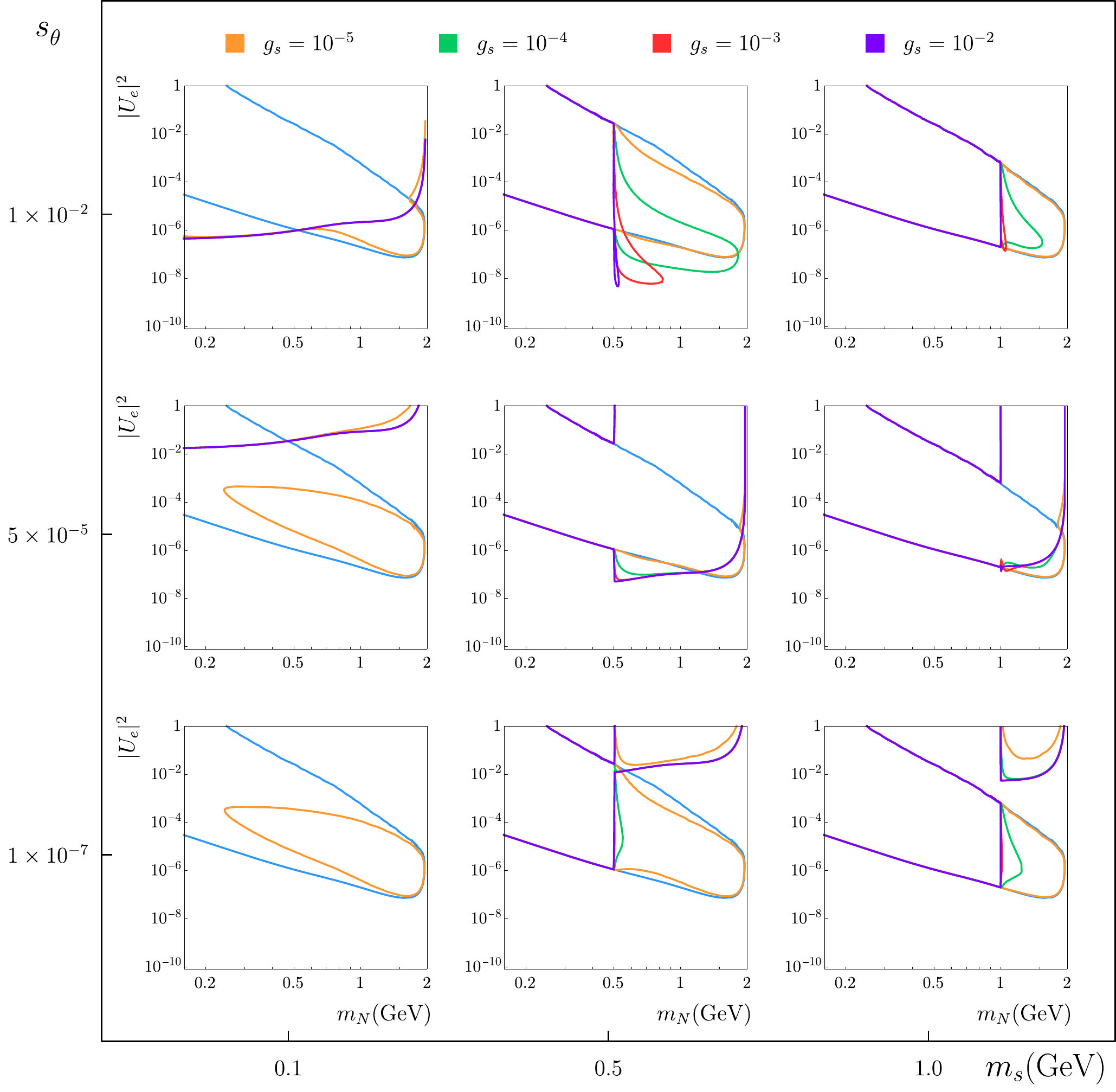}
    \caption{Recasted CHARM limits on $|U_e|^2$ versus HNL mass $m_N$, on a grid of scalar mixing versus mass. }
    \label{fig:CHARM-e}
\end{figure*}

\begin{figure*}[t]
    \centering
    \includegraphics[scale=0.75]{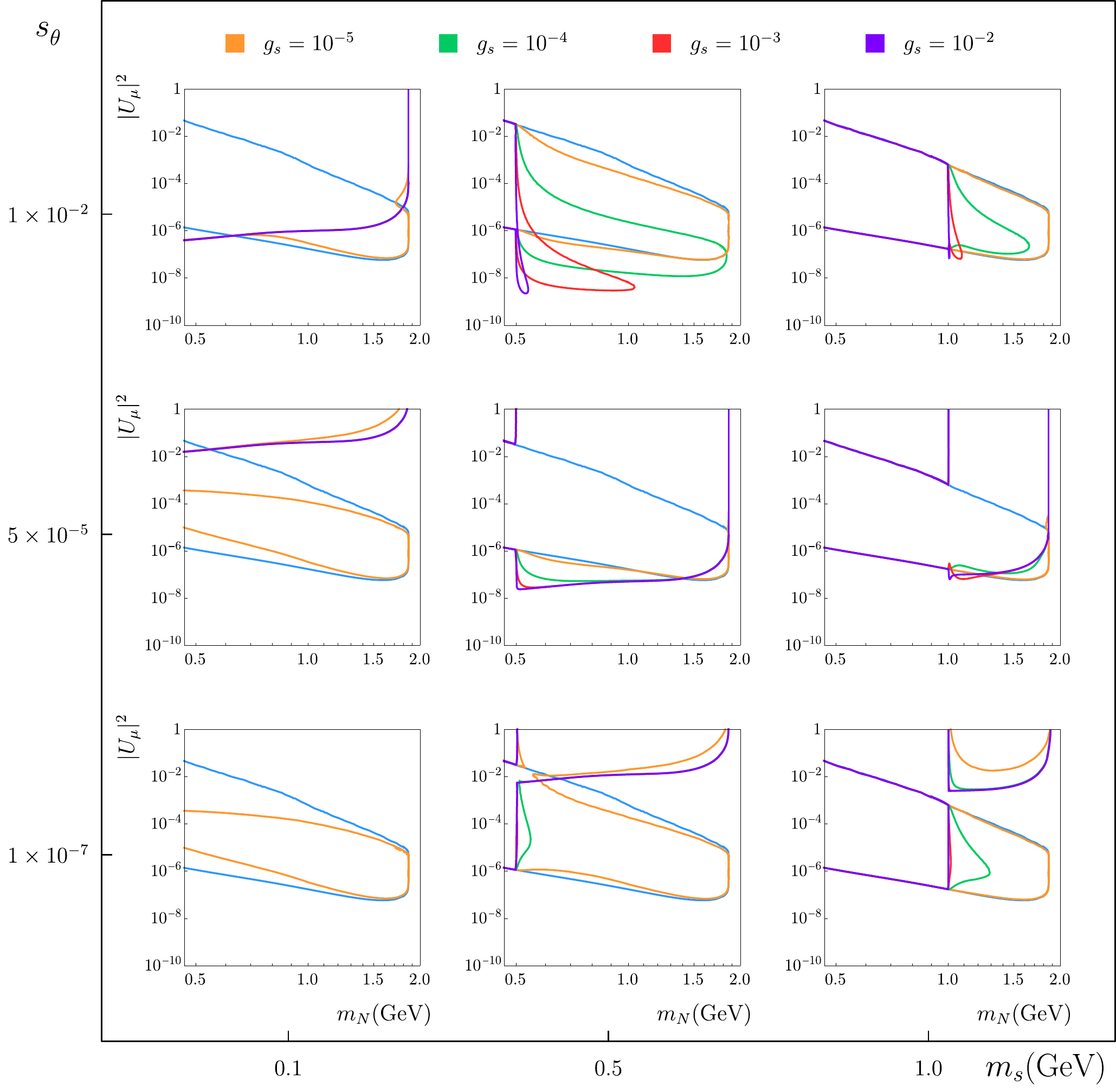}
    \caption{Like Fig.\ \ref{fig:CHARM-mu}, but constraining $|U_\mu|^2$ from the CHARM experiment. }
    \label{fig:CHARM-mu}
\end{figure*}

\begin{figure*}[t]
    \centering
    \includegraphics[scale=0.75]{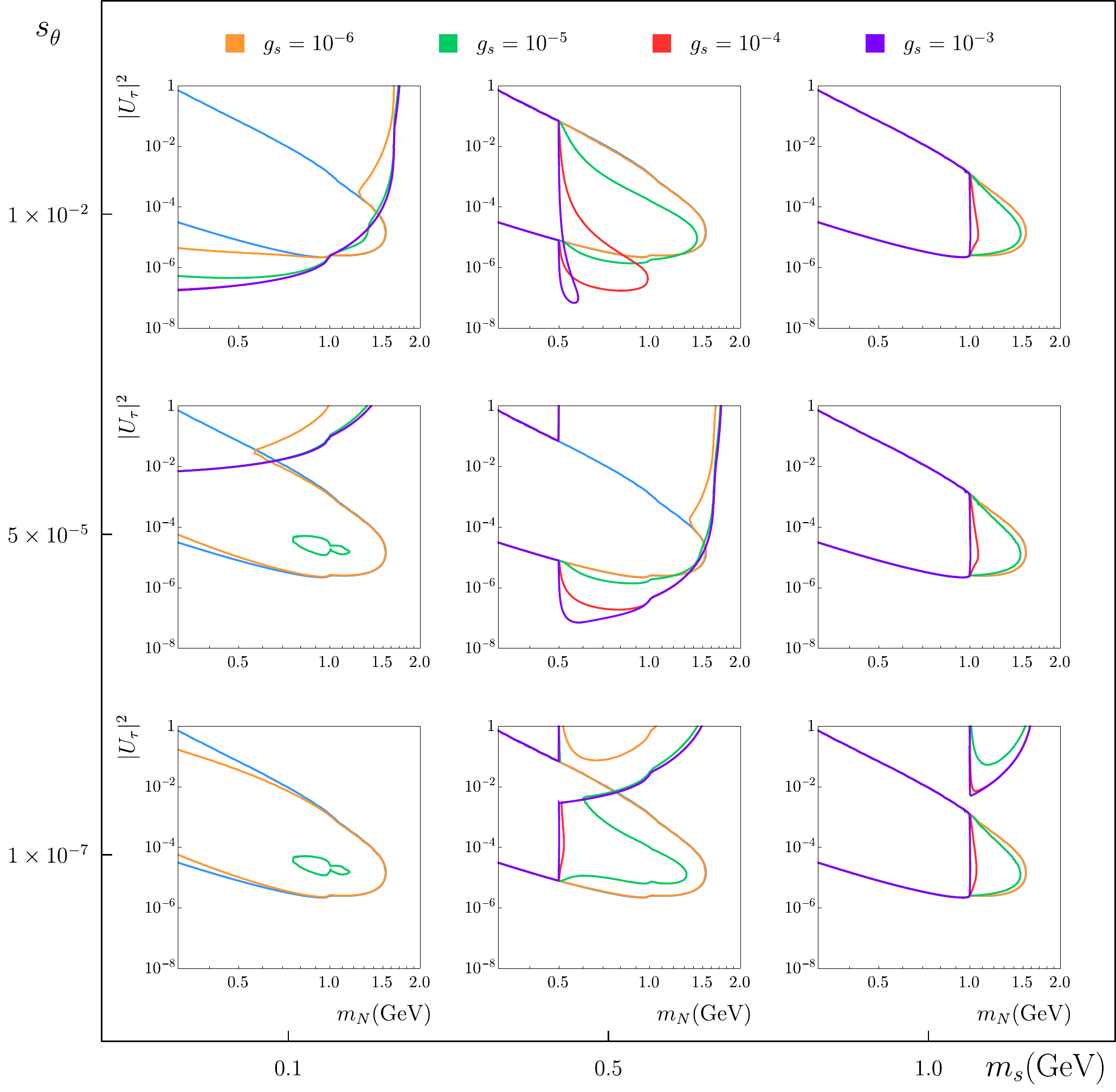}
    \caption{Recasted CHARM limits on $|U_\tau|^2$.}
    \label{fig:CHARM-tau}
\end{figure*}

Heavy neutral leptons can be produced in the semileptonic decays of $D^\pm$ and $D^0$, and in the leptonic decays of $D^\pm$. The CHARM Collaboration searched for HNLs in the mass range $0.5-1.8$ GeV \cite{charm}.  In subsequent analyses \cite{Gronau:1984ct,Boiarska:2021yho}, this range was expanded up to 2 GeV by considering
also the production of $D_s$, which we emulate here.

In the CHARM search, 400 GeV protons were stopped by a copper beam dump, producing $D$ mesons. These can decay into HNLs via mixing, whose subsequent decay produces one or two separate electromagnetic showers ($N \rightarrow e^+ e^- \nu_e$), two tracks ($N \rightarrow \mu^+ \mu^- \nu_\mu$), or one track and one electromagnetic shower ($N \rightarrow e^+ \mu^- \nu_e $ or $N \rightarrow \mu^+ e^- \nu_\mu$), in an empty decay region of length $D_L= 35$\,m and
cross-sectional area $3\times 3\,{\rm m}^2.$

The expected number of events is given by
\bea
N &=&  N_D \, \text{BR}(D\rightarrow N)\, \text{BR}(N \rightarrow \ell' \ell \nu_\ell) \, \mathcal{A} \nn \\
& \times & e^{-d/L_N} \left( 1- e^{-D_L/L_N} \right) \varepsilon
\label{eq-CHARM}
\eea
where $N_D$ is the number of $D$ mesons produced by protons in the dump, $\mathcal{A}$ is the acceptance factor (fraction of HNLs that enter the decay region), 
$d=480$\,m is the distance from the interaction point to the beginning of the decay region, $D_L$ is the length of the decay region, $L_N$ is the mean decay length of the HNL, and $\ell, \ell'=e,\mu$.  This formula is similar to
Eq.\ (\ref{BelleEq}) in the case where the reconstruction efficiency $\varepsilon$ is a constant, and the integration limits correspond to the boundaries of the detection region.\footnote{In the original CHARM analysis \cite{charm}, the distance from the interaction point to the beginning of the detector was ignored (taking $d=0$) incorrectly leading to exclusion of arbitrarily large mixing.}    The efficiency is $\varepsilon \sim 0.6$ for HNLs of mass $m_N\sim 1\,$GeV \cite{charm}.

To account for the singlet scalar decay channel, we modify Eq.\ (\ref{eq-CHARM})
similarly to the recasting of DELPHI, making the replacement
\bea
 \left( 1- e^{-D_L/L_N} \right) &\to& {\rm BR}_w \left( 1- e^{-D_L/L_N} \right)\\ &+& {\rm BR'}_s \left( 1- e^{-D_L/(L_N+L_s)} \right)\,,\nn
\eea 
where 
\begin{equation}
    {\rm BR'}_s = {\rm BR}_s \times \frac{{\rm BR}(s\to \ell \bar{\ell})}{{\rm BR}(N\to \ell'\ell \nu)},
\end{equation}
because the singlet $s$ not always decays into light lepton pairs. This is similar to the case of Belle, where BR$'_s=0$ since the singlet cannot decay into $\pi \ell$.

The CHARM limits are sensitive to lepton flavor, so we present respective constraints on $U_e$, $U_\mu$, $U_\tau$ for the cases of HNL coupling to a single
family, in Figures \ref{fig:CHARM-e}-\ref{fig:CHARM-tau}.  Although the original CHARM analysis did not include $U_\tau$ constraints, Ref.\ \cite{Boiarska:2021yho}
extended their results to do so by including neutral current contributions to the HNL decays, and we have done likewise.

Like the case of DELPHI, not only can constraints be weakened by the singlet decay channel, but in some regions of parameter space the signal can be enhanced by singlet decays
into $f\bar f$, leading to new excluded regions when 
singlet mixing times coupling ($\theta\,g_s$) is large enough. For example in the upper left plot of Fig.\ \ref{fig:CHARM-e}, the singlet decays within the detector
for most values of $g_s$, 
even when $|U_e|^2=1$, allowing exclusion of large mixing angles.  At larger $m_s$, the singlet starts to decay before reaching the experimental decay region, and the upper boundaries on the excluded regions reappear.  The bottom
right graph has a disconnected excluded region at large
$|U_e|^2$, due to the singlet decay length starting to fall within the detection region.  The constraints on $|U_\mu|^2$ shown in Fig.\ \ref{fig:CHARM-mu} are quite similar. Those on $|U_\tau|^2$ in Fig.\ \ref{fig:CHARM-tau} are qualitatively distinct, but display similar general features. We used the results of Ref.\ \cite{Boiarska:2021yho} to calibrate the sensitivity curves at low $|U_\tau|$.

\section{Relation to singlet scalar bounds}
\label{secIII}

\begin{figure*}[t]
    \centering
    \includegraphics[scale=1.1]{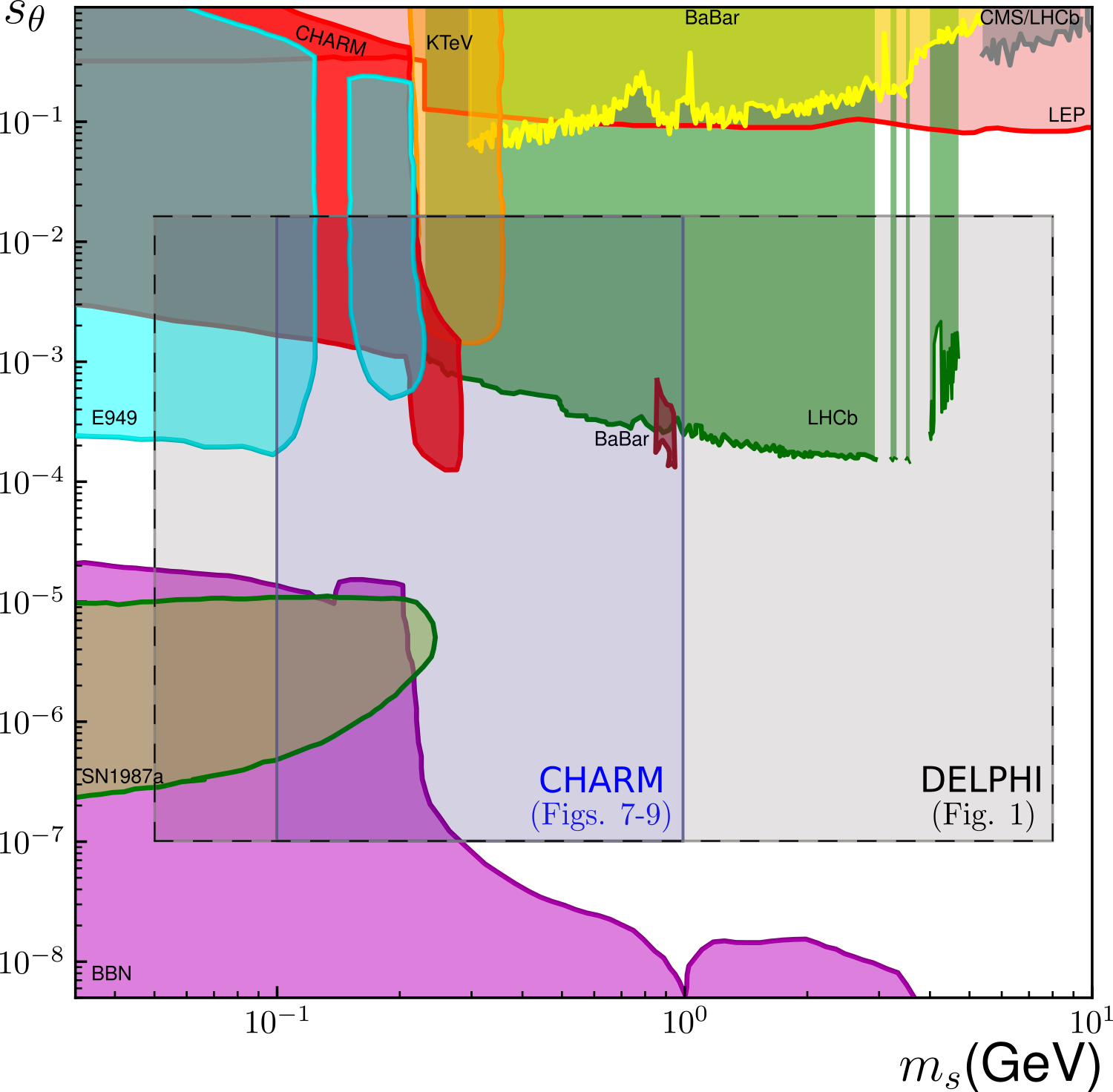}
    \caption{Current bounds on singlet scalar-Higgs boson mixing from LEP \cite{acciarri1996search}, CMS \cite{cms2012search}, LHCb \cite{LHCb:2018cjc}, CHARM \cite{charm}, E949 \cite{BNL-E949:2009dza}, BaBar \cite{PhysRevLett.107.221803,PhysRevD.88.071102,PhysRevLett.114.171801}, KTeV \cite{KTeV}, BBN \cite{Fradette:2017sdd}, and SN1987a \cite{Winkler:2018qyg}. Black and blue boxes show the $s_\theta-m_s$ parameter space used to get the results presented in Fig. \ref{fig:DELPHI} and Figs. \ref{fig:CHARM-e}-\ref{fig:CHARM-tau}, respectively. The results for Belle (see Fig. \ref{fig:Belle}) are independent from the singlet scalar-Higgs boson mixing $\theta$. Limits adopted from\cite{Winkler:2018qyg}.}
    \label{fig:singlet-constraints}
\end{figure*}

The Higgs mixing versus mass parameter space of the singlet, in which we
have displayed our recasted results for DELPHI and CHARM, is independently constrained by a variety of experiments or astrophysical considerations.
In Fig.\ \ref{fig:singlet-constraints} we have shown how the regions considered in our previous results compare with the previously constrained
parameter space.  There it can be seen that the largest mixing angle
$s_\theta = 0.01$ we considered is ruled out by beam dump experiments
or LHCb, except in the case of heavy singlets, $m_s \gtrsim 4\,$GeV.
Moreover for lighter singlets $m_s\lesssim 250\,$MeV, the region of small
mixing angles $s_\theta\sim 10^{-7}$ that we considered is excluded by
the effects of singlet decays on supernova 1987A or by BBN.  We have nevertheless included these regions in our analysis to give a complete picture of the qualitative trends.  It can be seen that our results overlap with a significant region of singlet parameters that is currently still open.

Just as the interactions of the singlet scalar with HNLs can alter the constraints on the HNL-active neutrino mixing angle,\footnote{even if the singlet does not decay into the signal that a given experiment ({\it e.g.,} Belle) is looking for}\ they can also affect the constraints on the singlet scalar-Higgs boson mixing. These changes in the limits come from modifications of the production and decays of the HNL and the singlet scalar.
The HNL width can be increased by the new decay channel $N\to s \nu$ or
by $N\to f\bar f \nu$ mediated by virtual $s$ exchange, competing with the
weak HNL 3-body decays. The singlet width can be increased by $s \to N \bar{\nu}$ or the analogous process with off-shell $N$ which decays weakly. The on-shell decays $N\to s \nu$ and $s \to N \bar{\nu}$ open when $m_N>m_s$ and $m_s>m_N$, respectively. Because these two regions of the full parameter space are mutually exclusive, our previous analysis is not affected by the new decay channels of the singlet.  Off-shell contributions would not modify our results either, since these mediate processes like $s \to \bar{\nu} N^* \to \bar{\nu} \ell W$, which is kinematically forbidden for the light scalars studied in this work. 

\section{Conclusions}
\label{secIV}
In this work we have estimated the changes to heavy neutral lepton mixing constraints due to their possible decays into a light singlet scalar and an active neutrino, for $m_N$ between 0.5 and $\sim 5$\,GeV.  One motivation for focusing on this mass range is the possibility that one generation of such HNLs could be the dark matter of the Universe
if they have sufficiently small mixing \cite{Cline:2020mdt}, while the other generations would be subject to the constraints investigated here.

It is possible that the limits derived here could be adapted to other qualitatively similar models.
For example, if HNLs couple to a light vector $Z'$,
which kinetically mixes with the standard model hypercharge, it would give rise to similar effects as we have studied, with $g_s$ representing the
new gauge coupling and $\theta y_f$ mapping onto
$\epsilon e$, where $y_f$ is the fermion Yukawa coupling (typically for the muon) and $\epsilon$
the kinetic mixing parameter.

Beyond the specific limits presented here, it may be that the general method
described could be useful for recasting other experimental constraints, especially in the case of older experiments where access to original data is not available, or Monte Carlo simulations would be difficult to carry out.

\bigskip

{\bf Acknowledgments}.  We are grateful to Dmitri Liventsev for extensive help in understanding and reproducing the Belle HNL limits, and to Gordan Krnjaic,  Maksym Ovchynnikov and Jonathan Rosner for very helpful correspondence. This work was supported by NSERC (Natural Sciences
and Engineering Research Council, Canada). GG acknowledges support from CNPq grant No.141699/2016-7 (Brasil), McGill Space Institute, and McGill Graduate $\&$ Postdoctoral Studies.

\begin{appendix}
\section{Methodology}

The first step in reproducing the observed limit of a given experiment is to compute the fractional
number $f$ of signal events, relative to decaying parent particles, on a grid in the 
$\log_{10}|U|^2-\log_{10}(m_N/\text{GeV})$ plane.  This is done without
necessarily knowing the overall normalization for the efficiency, or setting it to unity if the efficiency can be approximated as constant as a function of decay distance.
The normalized efficiency is inferred by plotting contours of $f$, and choosing that value of $f$ that best reproduces the published limit.  In addition, there may be other parameters that can be tuned in order to optimize the fit, namely the size $D_L$ of the decay region observed by the experiment. 

The fraction $f$ is generally determined by the product of
three probabilities,

\begin{equation}
f = {N_{E}\over N_{P}} =  \sum_{P,X,Y} \mathcal{P}_1(P,X)\, \mathcal{P}_2(Y)\, \mathcal{P}_3(P,X)\,,
\end{equation}
where $N_{E}$ is the number of observed events and $N_{P}$ is the number of  parent particles $P$ whose decays could produce HNLs. $N_{P}$ might be given or it might be computable from, for example, the number of protons on target and the production fractions \cite{Graverini:2133817,SHiP:2018xqw}.  The three probability factors are specified as follows.

$\mathcal{P}_1=\text{BR}(P \rightarrow N X)$ is the probability of HNL production for a given decay mode of $P$, where the HNL is accompanied by particles $X$. For example, for CHARM in the case of pure mixings with electron neutrinos, $X=e^+,\, e^+K^0,\,e^+K^{*0},\, e^+\pi^0$ with $P=D^+$. $X$ could be used to trigger for event candidates (Belle) or not (DELPHI, CHARM).

$\mathcal{P}_2=\text{BR}(N \rightarrow Y)$ is the probability for the HNL to decay into the signal being searched for. For DELPHI, $\mathcal{P}_2=1$  since all HNL decays compete with the decays of the Z bosons. In the case of Belle,  $\mathcal{P}_2=\text{BR}(N \rightarrow e \pi)$  for HNLs that mix only with $\nu_e$. (Recall that the signal events are then $ee\pi$ where the second lepton and the pion have opposite electric charge.) In contrast with DELPHI and Belle, the CHARM detector is far from the interaction point, so one must account for the fact that not all decay products of the HNLs travel toward the detector. The acceptance factor $\mathcal{A}$ quantifies this effect \cite{charm}. $\mathcal{A}$ generally depends on the mass of the HNL and the geometry of the experiment. In our analysis, it is taken as a free parameter to be fit by reproducing the sensitivity of the experiment, as further described below.

$\mathcal{P}_3$ is the probability of reconstructing the HNL from its decays when it was inside the detector. Its general form is
\begin{equation}
\mathcal{P}_3 = \int_d^{d+D_L} \frac{e^{-R/L_N}}{L_N}\,  \varepsilon (R,P,X)\, dR,
\label{A2}
\end{equation}
where $\varepsilon(R,P,X)$ is the reconstruction efficiency, which depends on the mass of the HNL, its production mode ($P,X$), and the distance $R$ from the interaction point to where it decays. 
The decay length $L_N$ depends on the HNL momentum, due to time dilation,
and this also depends on the 
 production modes, but it is more sensitive to the type of decay, {\it i.e.,} 2-body or 3-body, as we describe below.

In Eq.\ (\ref{A2}), $d$ is the distance from the interaction point to the beginning of the decay region inside the detector. For DELPHI where HNLs are created from the decays of Z bosons at rest, $d=0$. For Belle, since the background is higher near the interaction point, the experiment is insensitive to small-$R$ events, which are rejected by selection criteria
such that $\varepsilon\to 0$ at $R=0$.  Since the exact behavior of
$\varepsilon$ is uncertain near $R=0$, we take $d$ to be an undetermined
small cutoff to be fit by matching Belle constraints.
For CHARM the value $d=480$\,m is specified in their paper.

In order to calculate the mean decay length of the HNLs, $L_N=p_N/(m_N \Gamma_N)$, we calculated the momenta in the rest frame of the parent particles and boosted them to the laboratory frame, neglecting departures from the axis of the parent mesons. For 3-body decays $P \rightarrow N \ell x$ the maximum value of the momentum of the HNL is
\bea
    |\vec{p}_N^{\, (\text{max})}| &=& \frac{1}{2m_P} \bigg[ \bigg( m_P^2-(m_\ell + m_x + m_N)^2 \bigg) \nn \\
    &\times& \bigg(m_P^2-(m_\ell + m_x - m_N)^2 \bigg) \bigg]^{1/2}\!\!.
\eea
We used $|\vec{p}_N|=|\vec{p}_N^{\, (\text{max})}|/2$ before applying Lorentz transformations to the laboratory reference frame,  assuming $|\vec{p}_P| \approx$ 67 GeV, for $P=B^{\pm},D^{\pm},D^{0},$ and $D_s$ \cite{Gronau:1984ct}.

\begin{figure}[t]
\includegraphics[scale=0.8]{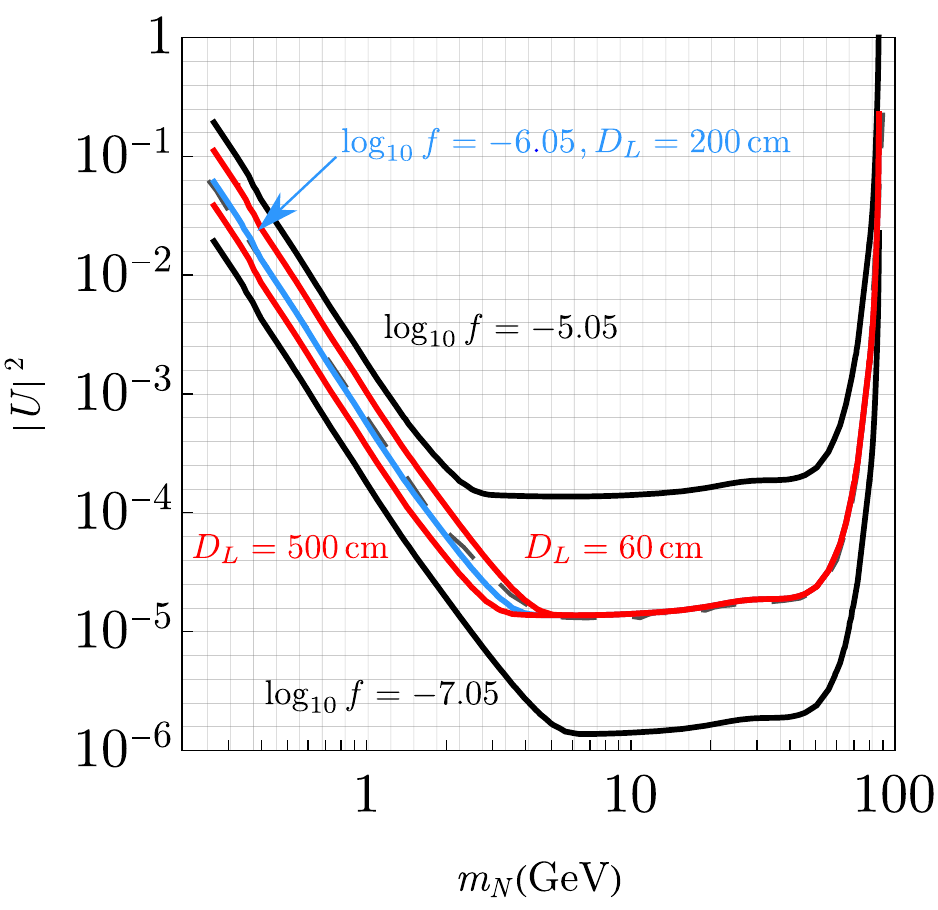}
\caption{Contours of detectable fraction $f$ of $Z$ bosons decaying into HNLs at DELPHI.  Black curves illustrate the
dependence on $f$, while red curves show the dependence on $D_L$, the assumed size of the decay detection length,
Eq.\ (\ref{eq-frac}).  Light blue curve is the best fit to the published limit (dashed curve).
}
\label{example-grid}
\end{figure}

Further details of this general procedure are next presented for the experiments of interest.  Fig.\ \ref{example-grid} shows contours of
the fraction $f$ of signal events for DELPHI.  The best-fit value is $\log_{10} f=-6.05$, indicating that the experiment was sensitive to one part in $\sim 10^6$ $Z$ bosons decaying into HNLs.  This is the correct order of magnitude, since
DELPHI produced $\sim10^6$ $Z$ bosons and observed one event, compared to an expected background of $0.8$ events.
The shape of the exclusion curve can be further tuned by
varying the length of the decay region $D_L$.  Choosing
$D_L=200$ cm provides the optimal fit, as illustrated by
the red curves in Fig.\ \ref{example-grid}.

For Belle, one must consider the dependence on the reconstruction efficiencies $\varepsilon (R)$ (see figure \ref{Belle-eff}). In this case, we have considered the limits integrating $R$ in Eq.\ (\ref{BelleEq}) to be from $d=1.75$\,cm to $d+D_L =$ 60\,cm. The curve that best reproduces the original constraint on $|U|^2$ is obtained with $\log_{10} f=-10.1$, regardless of the HNL flavor. With this choice, the total number of events is $N_E=2 \, N_{BB} \times 10^{-10.1} = 0.12$, where $N_{BB}=772 \times 10^6$ is the number of $B\bar{B}$ pairs at Belle, indicating the limits plotted Fig.\ \ref{results0} (center) are in accordance with the null results in the search for HNLs at Belle. For these calculations we have used updated formulas for $B$ meson branching ratios and HNL decay widths from Ref.\ \cite{Bondarenko:2018ptm}, after reproducing Belle's original limits based
on superseded branching ratios \cite{Gorbunov:2007ak}. The main difference between the updated
and original constraints is seen in the region near $m_N \gtrsim$ 2\,GeV because of revisions in the branching ratios for the $B\to \rho N \ell$
and $B\to \pi N \ell$ production modes.

For CHARM, we integrate from $d=480$\,m  to $d+D_L = 515\,$m  \cite{Gronau:1984ct}. The excluded region in the $|U|^2-m_N$ plane is located to the left of the light blue curve in Fig. \ref{results0} (Right). 
In order to interpret $f$ as the fraction of events for a null search for HNLs in the CHARM experiment, we take the efficiencies to be $60\%$ as reported by the CHARM collaboration for $m_N=$1 GeV \cite{charm}, and the
acceptance factor is $\mathcal{A}\cong 10^{-3}$ (see Fig.\ 7 of Ref.\ \cite{Boiarska:2021yho}).  We then fix $\log_{10} f=-15.6$ to match the original limit, giving the number of events $N_E = N_D \times f = 2.24 < 2.3$ at 90$\%$ C.L.. The number of $D$ mesons was determined using $N_{D_i}=N_{POT} \times \chi_{c\bar{c}} \times f_{c\rightarrow D_i}$, $N_{POT}=2.4 \times 10^{18}$, $\chi_{c\bar{c}} \approx 4 \times 10^{-3}$ for a 400 GeV proton beam  \cite{CERN-SHiP-NOTE-2015-009}, $f_{c\rightarrow {D^+}}=0.207$,  $f_{c\rightarrow {D^0}}=0.632$, and $f_{c\rightarrow {D_s}}=0.088$  \cite{Graverini:2133817,SHiP:2018xqw}.
 (``POT" denotes ``protons on target.")

\end{appendix}

\bibliographystyle{utphys}
\bibliography{refs.bib}
\end{document}